\documentclass[10pt]{article}

\usepackage[dvips]{graphicx}

\newcommand{\qed}{\nobreak \ifvmode \relax \else
      \ifdim\lastskip<1.5em \hskip-\lastskip
      \hskip1.5em plus0em minus0.5em \fi \nobreak
      \vrule height0.75em width0.5em depth0.25em\fi}

\title{Generalized Leapfrogging Samplesort: A Class of $O(n \log^2 n)$ Worst-Case Complexity and $O(n \log n)$ Average-Case Complexity Sorting Algorithms}

\author{
Eliezer A. Albacea \\
Institute of Computer Science \\
University of the Philippines Los Ba\~{n}os \\
College, Laguna, Philippines \\
eaalbacea@up.edu.ph}

\begin{document}
\maketitle

\begin{abstract}
The original Leapfrogging Samplesort operates on a sorted sample of size $s$ and an unsorted part of size $s+1$. We generalize 
this to a sorted sample of size $s$ and an unsorted part of size $(2^k-1)(s+1)$, where $k = O(1)$. We present a practical implementation 
of this class of algorithms and we show that the worst-case complexity is $O(n \log^2 n)$ and the average-case complexity is $O(n \log n)$. \\ \\
Keywords: Samplesort, Quicksort, Leapfrogging Samplesort, sorting, analysis of algorithms.
\end{abstract}

\section{Introduction}
Samplesort was shown by Frazer and McKellar \cite{fra} to be a sorting algorithm that has a potential of competing with 
Quicksort \cite{hoa2} in terms of average running time. In fact, it was shown in Frazer and McKellar \cite{fra} that Samplesort average running 
time slowly approaches the information-theoretic lower bound. Apers \cite{ape}, on the other hand, improved Samplesort by introducing 
Recursive Samplesort. The idea is to make use of Samplesort itself, instead of Quicksort, in sorting the sample. The expected 
number of comparisons of Recursive Samplesort was shown in Apers \cite{ape} to be close to the information-theoretic lower bound. 
Another implementation of Samplesort was given by Peters and Kritzinger \cite{pet}. Unfortunately, not one of these implementations 
of Samplesort can be considered practical. The implementation of Peters and Kritzinger \cite{pet}, for example, uses temporary storage
 locations for storing the sample which eventually are used to store pointers to positions in the array bounded by the sample. 
The implementation of Apers \cite{ape}, on the other hand, uses a stack to store pointers to positions in the array that are bounded 
by the sample. All the implementations run in $O(n^2)$ worst-case time. 

In 1995, Albacea \cite{alb1} reported the algorithm Leapfrogging Samplesort which is a practical implementation of Samplesort. 
The algorithm has a worst-case complexity of $O(n \log^2 n)$ \footnote{All logarithms in this paper are to base 2, except when it is explicitly 
stated.} and an average-case complexity of $O(n \log n)$. Albacea \cite{alb2} estimated the exact average-case complexity to a value that is 
very near the information-theoretic lower bound. Chen \cite{chen}, in 2006, proposed the algorithm Full Sample sort whose worst-case complexity 
is $O(n \log^2 n)$ and whose average-case complexity is $O(n \log n)$.

In this paper, we introduce a generalization of Leapfrogging Samplesort, where we have a sample of size $s$ and an unsorted part of size $(2^k-1)(s+1)$ where 
$k = O(1)$. When $k > O(1)$, say $O(\log n)$, the algorithm reduces to Quicksort. The generalized Leapfrogging Samplesort has a worst-case complexity of $O(n \log^2 n)$ and an average-case complexity of $O(n \log n)$. Thus,this class 
of algorithms extends the number of practical algorithms whose worst-case complexity is $O(n \log^2 n)$ and whose average-case complexity is $O(n \log n)$. 
The author is aware of only two such algorithms in this class of algorithms, Leapfrogging Samplesort by Albacea \cite{alb1} and Full Sample Sort by Chen \cite{chen}.

\section{Generalized Leapfrogging Samplesort}
The original Leapfrogging Samplesort involves in each stage of the sorting process the first $(2s+1)$ elements of
the sequence, where the first $s$ elements are already sorted and the next $(s+1)$ elements are to be partitioned and 
sorted using the sorted $s$ elements as the sample. 

The algorithm starts with the leftmost element as a sorted sample of size $1$ that is used to partition the next $2$ elements, eventually producing a sorted sequence of size $3$. The sorted sequence 
of size $3$ is used as a sample to partition the next $4$ elements, eventually producing a sorted sequence of size $7$. The 
sorted sequence of size $7$ is used to partition the next $8$ elements, eventually producing a sorted sequence of size $15$. 
The process is repeated until the whole sequence is sorted.

Given a sequence prefixed by a sample of size $s$ and an unsorted part whose size is at most $s+1$, an outline of the algorithm for partitioning the unsorted part using the sorted sample is as follows:

Step $1$: Let $m$ be the middle element of the sorted sample and the group of elements to the left is the left subsample and the group of elements 
to the right is the right subsample. Using $m$ as a pivot element, we partition the unsorted part thereby producing two partitions, namely: the left 
partition (elements which are less than $m$)\footnote{Without loss of generality, we assume that the elements of the sequence are distinct.} and the 
right partition (elements which are greater than $m$). Then, $m$ and the right subsample are moved to the left of the right partition and the left 
partition, is moved to the right of the left subsample. This step will produce two subsequences where each subsequence is prefixed by a sorted sample. 

Step $2$: Recursively apply Step $1$ until the size of the sorted sample is equal to $1$ on the two sequences produced in Step $1$.

If after the partitioning process, a partition whose size is greater than $1$ is produced, then such partition is sorted by Leapfrogging Samplesort itself.

Table \ref{table1} illustrates the sizes of the sorted and unsorted parts using the ratio $s:s+1$.

\begin{table}[ht]
\begin{center}
\begin{tabular}{|r|r|}
\hline
sorted sample&unsorted part\\
\hline
1&2\\
\hline
3&4\\
\hline
7&8\\
\hline
15&16\\
\hline
31&32\\
\hline
63&64\\
\hline
127&128\\
\hline
255&256\\
\hline
...&...\\
\hline
\end{tabular}
\end{center}
\caption{Sizes of the sorted and unsorted parts using the ratio $s:s+1$.}
\label{table1}
\end{table}

A generalization of this is obtained by reducing the ratio between the sizes of the sorted sample and the unsorted part. One such 
class of ratios is the ratio defined by $s:(2^k-1)(s+1)$ where $k = O(1)$, $s$ is the size of the sorted sample and $(2^k-1)(s+1)$ 
is the size of te unsorted part. Of course with $k = 1$, this reduces to the original Leapfrogging Samplesort. Table \ref{table2} 
illustrates the sizes of the sorted and unsorted parts for $k = 2$ to $4$.

\begin{table}[ht]
\begin{center}
\begin{tabular}{|r|r|r|r|r|r|}
\hline
\multicolumn{2}{|c|}{$k=2$} & \multicolumn{2}{c|}{$k=3$} & \multicolumn{2}{c|}{$k=4$} \\
\hline
sorted sample & unsorted part & sorted sample & unsorted part & sorted sample & unsorted part\\
\hline
$s$ & $3(s+1)$ & $s$ & $7(s+1)$ & $s$ & $15(s+1)$ \\
\hline
$1$ & $6$ & $1$ & $14$ & $1$ & $30$ \\
\hline
$7$ & $24$ & $15$ & $112$ & $31$ & $480$ \\
\hline
$31$ & $96$ & $127$ & $896$ & $511$ & $7680$ \\
\hline
$127$ & $384$ & $1023$ & $7168$ & $8191$ & $122880$ \\
\hline
$511$ & $1536$ & $8191$ & $57344$ & $131071$ & $1966080$ \\
\hline
... & ... & ... & ... & ... & ... \\
\hline
\end{tabular}
\end{center}
\caption{Sizes of sorted and unsorted parts using the ratio $s:(2^k-1)(s+1)$ where $k = 2$ to $4$.}
\label{table2}
\end{table}

A practical implementation of the generalized Leapfrogging Samplesort is given below:

\begin{verbatim}
        void LFSamplesort(int first, int last)
        {
                int s;
                int r;
                if (last > first) {
                        s = 1;
                        r = M*(s+1);	
                        while (s <= (last-first+1-r)) {
                                Leapfrog(first, first+s-1, first+s+r-1);
                                s = s+r;
                                r = M*(s+1);
                                }
                        Leapfrog(first, first+s-1, last);
                        }
        }

\end{verbatim}

The constant $M = 2^k-1$.

\begin{verbatim}
        void Leapfrog(int s1, int ss, int u)
        {
                int i,j,k, sm, v,t;
                if (s1 > ss) LFSamplesort(ss+1, u);
                else
                if (u > ss) {
                        sm = (s1+ss) / 2;
                        /* Partition */
                        v = A[sm];
                        j = ss;
                        for(i=ss+1; i <= u; i++) {
                                if (A[i] < v) {
                                        j++;	
                                        t = A[j];
                                        A[j] = A[i];
                                        A[i] = t;
                                        }
                                }
                        /* Move Sample */
                        if (j > ss) {
                                for (k=j, i=ss; i >= sm; k- -, i- -) {
                                        t = A[i];
                                        A[i] = A[k];
                                        A[k] = t;
                                        }
                                }
                        Leapfrog(s1, sm-1,sm+j-ss-1);
                        Leapfrog(sm+j-ss+1, j, u);
                        }
         }
\end{verbatim}

The code above of the generalized Leapfrogging Samplesort is similar to the code of the Leapfrogging Samplesort given in 
Albacea \cite{alb2}, except for a minor difference. Specifically, the difference between the two codes is the introduction 
of constant $M$ to the code of the generalized Leapfrogging Samplesort.

\section{Worst-Case Analysis}
The operation that dominates the execution of the algorithm is the comparison operation. Hence, our analysis will be 
in terms of number of comparisons involved in the algorithm. We refer to the number of comparisons involved in the 
algorithm as the cost of the algorithm.

The worst case is exhibited when the values of the sample are all less than or all greater than the unsorted 
elements every time the unsorted portion is partitioned using the elements of the sample. Without loss of generality, 
we assume $n = s+(2^k-1)(s+1)$. With this value of $n$, we obtain a worst-case complexity of:
\[W(n) = W(s)+W((2^k-1)(s+1))+(2^k-1)(s+1)\log(s+1)\]
where $W(s)$ is the cost of applying Leapfrogging Samplesort on the
sample of size $s$, $W((2^k-1)(s+1))$ is the cost of sorting using 
Leapfrogging Samplesort the unsorted
sequence of size $(2^k-1)(s+1)$ which remains unsorted after the partitioning
process, and $(2^k-1)(s+1)\log(s+1)$ is the cost of partitioning the 
unsorted sequence of size $(2^k-1)(s+1)$ using a sorted sample of size $s$.

When $k=1$, given 
\[n = s+(s+1)\]
\[s = \frac{n-1}{2}\]
we obtain the recurrence relation
\[W(n) = W(\frac{n-1}{2})+W(\frac{n+1}{2})+(\frac{n+1}{2})\log(\frac{n+1}{2})\]
\[W(n) = O(n \log^2 n).\]

When $k=2$, similarly, given
\[n = s+3(s+1)\]
\[s = \frac{n-3}{4}\]
we obtain the recurrence relation
\[W(n) = W(\frac{n-3}{4})+W(\frac{3(n+1)}{4})+(\frac{3(n+1)}{4})\log(\frac{n+1}{4})\]
\[W(n) = O(n \log n \log_\frac{4}{3} n)\]
\[W(n) = O(n \log^2 n).\]

For any integer $k > 0$, $k = O(1)$, given
\[n = s+(2^k-1)(s+1)\]
\[s = \frac{n-(2^k-1)}{2^k}\]
we produce the recurrence relation
\[W(n) = W(\frac{n-(2^k-1)}{2^k}) + W(\frac{(2^k-1)(n+1)}{2^k})+(\frac{(2^k-1)(n+1)}{2^k})\log(\frac{n+1}{2^k})\]
\[W(n) = O(n \log n \log_\frac{2^k}{2^k-1} n)\]
\[W(n) = O(n \log^2 n).\]

\section{Average-Case Analysis}
Without loss of generality, we assume $n = s+(2^k-1)(s+1)$. The average-case complexity of the algorithm is given by the recurrence relation 

\[A(n) = A(s)+(2^k-1)(s+1)\log(s+1)+(s+1)A(2^k-1)\]
where $A(s)$ is the average cost of sorting the sample of size $s$, $(2^k-1)(s+1)\log(s+1)$ is the cost of partitioning the unsorted part of 
size $(2^k-1)(s+1)$ using the sorted sample of size $s$. The idea is that the middle element of the sample will be used as a pivot element in partitioning 
the unsorted part of size $(2^k-1)(s+1)$. Using Lemma 1 of Frazer and McKellar\cite{fra}, the expected size of each of the 2 partitions is $\frac{1}{2}$ the size of the 
unsorted part, provided the pivot element is a random sample of size 1 from the set composed of the pivot element and elements of the unsorted part. Then, using the first 
quarter and the third quarter elements of the sorted sample as pivot elements, we split each partition into 2 more partitions. We continue doing this for $\log(s+1)$ steps. 
This will produce $s+1$ partitions where the expected size of each partition is $2^k-1$. Hence, the cost of sorting the $s+1$ partitions is $(s+1)A(2^k-1)$. Given 

\[n = s+(2^k-1)(s+1)\]
\[s = \frac{n-(2^k-1)}{2^k}\]
will produce the recurrence relation
\[A(n) = A(\frac{n-(2^k-1)}{2^k})+(2^k-1)(\frac{n+1}{2^k}\log\frac{n+1}{2^k})+\frac{n+1}{2^k}O(1)\]
\[A(n) = O(n \log n)\]
where $A(2^k-1) = O(1)$, when $k = O(1)$

\section{Conclusions}

We have presented a practical implementation of a generalized Leapfrogging Samplesort and analyzed its worst-case complexity and average-case complexity. It was shown that the 
worst-case complexity is $O(n \log^2 n)$ and the average-case complexity is $O(n \log n)$. Thus, extending the number of practical algorithms whose worst-case complexity is 
$O(n \log^2 n)$ and whose average-case complexity is $O(n \log n)$. What remains open is the computation of the exact average-case complexity of the generalized Leapfrogging Samplesort.

\end{document}